\documentclass[sn-mathphys,Numbered,onecolumn]{sn-jnl}

\usepackage{graphicx}%
\usepackage{multirow}%
\usepackage{amsmath,amssymb,amsfonts}%
\usepackage{amsthm}%
\usepackage{mathrsfs}%
\usepackage[title]{appendix}%
\usepackage{xcolor}%
\usepackage{textcomp}%
\usepackage{manyfoot}%
\usepackage{booktabs}%
\usepackage{algorithm}%
\usepackage{algorithmicx}%
\usepackage{algpseudocode}%
\usepackage{listings}%
\usepackage{dirtytalk}%
\usepackage{ulem}%

\newcommand\hide[1]{{}}

\newcommand{\JSD}[2]{D_{\mathrm{JS}}\left(#1,#2\right)}
\newcommand{\JSDhat}[2]{\hat{D}_{\mathrm{JS}}\left(#1,#2\right)}

\newcommand{\KL}[2]{D_{\mathrm{KL}}\left(#1\middle\|#2\right)}
\newcommand{\KLhat}[2]{\hat{D}_{\mathrm{KL}}\left(#1\middle\|#2\right)}
\newcommand{\Df}[2]{D_{f}\left(#1\middle\|#2\right)}

\newcommand{\TV}[2]{D_{\mathrm{TV}}\left(#1||#2\right)}
\newcommand{\D}[2]{D\left(#1 || #2\right)}

\newcommand{\expect}[1]{\mathbb{E}\left[#1\right]}
\newcommand{\expwrt}[2]{\mathbb{E}_#1\left[#2\right]}

\newcommand{\Lbar}{\bar{\mathcal{L}}}
\newcommand{\Lhat}{\hat{\mathcal{L}}}
\newcommand{\phimin}{\phi_\mathrm{min}}

\theoremstyle{thmstyleone}%
\theoremstyle{thmstyletwo}%

\theoremstyle{thmstylethree}%

\begin{document}

\title[Statistical divergences in high-dimensional hypothesis testing and a modern technique for estimating them]{Statistical divergences in high-dimensional hypothesis testing and a modern technique for estimating them}


\author[]{\fnm{Jeremy J.H.} \sur{Wilkinson}}\email{jjhw3@cam.ac.uk}

\author[]{\fnm{Christopher G.} \sur{Lester}}\email{cgl20@cam.ac.uk}


\affil[]{\orgdiv{Department of Physics}, \orgname{Cavendish Laboratory}, \orgaddress{\street{JJ Thomson Avenue}, \city{Cambridge}, \postcode{CB3 0HE}, \state{Cambridgeshire}, \country{United Kingdom}}}

\abstract{Hypothesis testing in high dimensional data is a notoriously difficult problem without direct access to competing models' likelihood functions. This paper argues that statistical divergences can be used to quantify the difference between the population distributions of observed data and competing models, justifying their use as the basis of a hypothesis test.
We go on to point out how modern techniques for functional optimization let us estimate many divergences, without the need for population likelihood functions, using samples from two distributions alone. We use a physics-based example to show how the proposed two-sample test can be implemented in practice, and discuss the necessary steps required to mature the ideas presented into an experimental framework. The code used has been made available for others to use.}





\maketitle


\section*{Introduction}

Between the axioms of Bayesian probability theory and the Neyman–Pearson lemma, log-likelihood ratio based tests are generally accepted as the logical way of deciding between competing hypotheses.
However, typical likelihood-ratio based tests rely on two assumptions which may not be fulfilled in practice:
\begin{enumerate}
    \item Given arbitrary data, $\{x_i\}_{i=1}^N$, we have access to the likelihood, $p(x|H)$, for each data point and competing hypothesis $H$.
    \item The true underlying process explaining our data is in fact within the set of models we consider. In our own words, we must be working in a \emph{complete model space}.
\end{enumerate}
In the absence of the first condition, it is clear that a parametric test simply cannot be performed.
This situation readily arises in experiments with a moderately complex measurement apparatus, since the model of one's detector response, stacked on top of the physical model of interest, typically won't admit a computable likelihood function.
A number of work arounds are used to bypass the need for direct access to the likelihood functions.
A common approach relies on Monte-Carlo simulations to produce simulated data that may be used to fit functional forms approximating the model's likelihood function or used to perform a binned likelihood test \cite{Cowan2011,LHCB}.
These approaches have seen phenomenal success, but require significant data to provide a reasonable approximation of each competing model's likelihood function.
Since effective sampling density falls exponentially as a function of the number of dimensions in the data, such techniques are difficult to apply to high dimensional data.
One is typically forced to marginalise over a significant number of the measured dimensions in order to obtain sufficiently high data-density to apply the aforementioned techniques.
Unfortunately, marginalizing data comes with the risk of significantly reducing the sensitivity of the test, and it is not always clear which dimensions should be marginalized over.
Performing a marginalised analysis of one's data, finding nothing of interest, and performing the analysis again but marginalising over a different set of variables is precisely what the look-elsewhere effect warns of and is bound to produce unreliable conclusions.
We believe that as scientific endeavours increase in complexity over time, we will inevitably have to embrace forms of hypothesis testing which do not rely on direct access to likelihood functions while retaining as much sensitivity as possible.

The philosophy of this paper revolves around the principal that one should favour models which predict a distribution of data most similar to the distributions we observe.
There are many ways of measuring the similarity of two distributions, $p$ and $q$, but we restrict our discussions to statistical divergences. That is, functions $D$ satisfying
\begin{enumerate}
    \item $\D{p}{q} \geq 0$
    \item $\D{p}{q} = 0 \iff p = q.$
\end{enumerate}
A number of divergences have become very well-known. These include,
\begin{enumerate}
    \item the Kullback-Leibler (KL) divergence, $\KL{p}{q} := \expwrt{q}{\frac{p(x)}{q(x)}\ln{\frac{p(x)}{q(x)}}}$,
    \item the chi-squared, $\chi^2\left(p||q\right) := \expwrt{q}{\frac{1}{q(x)}\left(p(x)-q(x)\right)^2}$,
    \item the Total Variational distance $\TV{p}{q} = \frac{1}{2} \expwrt{q}{\left|\frac{p(x)}{q(x)} - 1\right|}$,
    \item and the Jensen-Shannon divergence, $$\JSD{p}{q} = \frac{1}{2}\left(\vphantom{\int}\KL{\,p\,}{\ \tfrac{1}{2}(p+q)\,} + \KL{\,q\,}{\ \tfrac{1}{2}(p+q)\,}\right).$$
\end{enumerate}
Since the proliferation of back-propagation as a tool for functional optimization, a number of authors have pointed out the potential for machine learning as a tool for estimating the divergence between two distributions using nothing more than a set of samples from each. 
No direct access to the likelihood function required.
We wonder whether the significance of this result has been overlooked by the scientific community, and whether we may turn these results into a tool for inference which is maximally sensitive to differences between models and data while remaining effective and practical in arbitrarily high dimensions.

Before diving into technical details, we believe it is important to confront one more philosophical point. 
The Bayesian school of thought is founded upon a \emph{unique} set of logically consistent rules for combining degrees of belief in a set of statements. 
Similarly, from the frequentest perspective, the Neyman-Pearson lemma proves that the log-likelihood ratio is the uniquely most powerful test statistic (the UMP) for deciding between two hypothesis $H_0$ and $H_1$. 
Does this immediately doom a divergence-based hypothesis test to be inherently sub-optimal? The short answer is no.
The reason for this is a subtle consequence of the inevitable violation of the second condition described above.
Even if you're not convinced by the argument that `all models are approximate', one can never know one's detector response to infinite precision, nor can one control for all variations in an experiment's initial setup and external influences.
Assumptions made about these factors are implicitly built into each and every hypothesis and have tangible effects on their likelihood functions.
Therefore we are always necessarily working in an incomplete model space.
The proof of the Neyman-Pearson lemma relies on the assumption that either $H_0$ or $H_1$ provides the true explanation of our data and therefore predicts the observed data distribution exactly. Likewise, if one is to be strictly honest within the Bayesian framework, how can one assign a non-zero degree of belief to any hypothesis in an incomplete model space? 
Ignoring this fact and simply cranking the mathematical handle requires the concession that the quantities we are calculating are not degrees of belief, but rather something else, to which the rules of Bayesian inference need not apply.
Then again no amount of pessimism about the foundations of these techniques can refute their incontrovertible record of success in practice.
A reasonable response might point out that clearly these tests cannot pick the true underlying hypothesis out of an incomplete model space (by definition), but evidence suggests that they must be selecting the model which is, by at least some definition, most similar to the data.
Indeed, this point of view forms the core of this paper, and it is surprisingly easy to prove that these tests implicitly use the KL divergence as their measure of `similarity'. To understand why, consider a standard log-likelihood ratio test which rejects the null hypothesis if the test statistic,
$$
\hat{t} = \sum_{i=1}^N \log \left(\frac{L(x_i | H_1)}{L(x_i | H_0)}\right),
$$
is greater than some decision threshold value $\alpha$. In a complete model space, the Neyman-Pearson lemma assures us that $\hat{t}$ is a UMP, but how should we understand this procedure in an incomplete model space in which the underlying data is in fact explained by an unknown alternative model $H_T$? In this case it is instructive to re-arrange $\hat{t}$,
$$
\hat{t} = \sum_{i=1}^N \log \left(L(x_i | H_1)\right) -\sum_{i=1}^N \log\left(L(x_i | H_0)\right)
$$
$$
= N \left(\frac{1}{N} \sum_{i=1}^N \log \left(\frac{L(x_i | H_T)}{L(x_i | H_0)}\right) - \frac{1}{N} \sum_{i=1}^N \log\left(\frac{L(x_i | H_T)}{L(x_i | H_1)}\right)\right).
$$
Since the observed data points are themselves drawn from the underlying distribution $L(x|H_T)$, $\hat{t}$ is in fact an unbiased estimator for the difference between KL divergence of the competing hypotheses and the true data distribution,
$$
\expect{\hat{t}} = N \left(\expwrt{{H_T}}{\log \left(\frac{L(x | H_T)}{L(x | H_0)}\right)} - \expwrt{{H_T}}{\log\left(\frac{L(x | H_T)}{L(x | H_1)}\right)}\right)
$$
$$
= N \left(\KL{L(\cdot|H_T)}{L(\cdot|H_0)} - \KL{L(\cdot|H_T)}{L(\cdot|H_1)}\right).
$$
As a result, regardless of the value of $\alpha$, as $N \rightarrow \infty$ the law of large numbers assures us that the standard log-likelihood ratio test picks out the hypothesis with the smallest KL-divergence between the predicted data distribution and the true underlying data distribution.
Remarkably, not only are the concerns that divergence-based hypothesis tests might be provably sub-optimal unfounded, we have in fact been performing divergence-based hypothesis testing all along. A very similar argument may be applied to the Bayesian framework in an incomplete model space to reach the same conclusion.

Once you have let go of the idea that the KL divergence is somehow the \emph{uniquely correct} way of deciding between competing hypotheses in an incomplete model space, wherein the theorems of optimality do not apply, one may consider many interesting alternative divergence-based techniques.\footnote{Indeed,  recurring recognition of the need to have good ways of testing whether samples are (or are not) drawn from the same population has motivated  related approaches from others in physics.  This is particularly so in cases where samples live in high-dimensional spaces and/or where underlying likelihoods are not computable. See for example Ref.~\cite{Barter:2018xbc}. }


\section{A simple example using binary classification} \label{sec:jsd}

We introduce the relationship between divergences and machine learning using the familiar problem of binary classification and a trick first written about, as far as we know, in the literature on Generative Adversarial Networks \cite{GAN}. A binary classifier, $\phi$, used to discriminate samples from two categories, $p$ and $q$, is typically trained to minimize a Binary Cross Entropy (BCE) loss function,
\begin{equation}
    \label{eq:loss}
    \Lhat[\phi] := -\frac{1}{N} \sum_{i=1}^N \left( 
    \vphantom{\int} 
    l_i \log(\phi(x_i)) + (1 - l_i) \log(1 - \phi(x_i)) \right),
\end{equation}
where each data point $x_i$ is paired with a corresponding label
$$
l_i = \begin{cases}
    0 & \text{if }x_i \text{ was drawn from } p, \\
    1 & \text{if }x_i \text{ was drawn from } q.
\end{cases}
$$
Assuming an equal number of samples from both categories, the expected value of the loss for fixed $\phi$ is given by
$$
\Lbar[\phi] := \expect{\Lhat[\phi]} = - \int dx \left(
 \vphantom{\int} 
 \frac{1}{2}p(x)\log(\phi(x)) + \
 \frac{1}{2}q(x) \log(1 - \phi(x))\right).
$$
A functional variation with respect to $\phi$ readily shows the well known result that the expected loss is minimized at $\phimin(x) = L(p|x) = \frac{p(x)}{p(x)+q(x)}$, the likelihood that the given sample $x$ was drawn from category $p$.
We refer to this critical function as the optimal classifier for $p$ and $q$\footnote{Since for any $x$, $\phimin(x) \in \left[0, 1\right]$, the classifier's output may be constrained to fall in $\left[0, 1\right]$ without any loss of generality.}.
Substituting $\phimin$ into $\Lbar$ demonstrates that the minimum expected loss, attained by the optimal classifier, is related to the Jensen-Shannon divergence between the category distributions,
\begin{align*}
\Lbar[\phimin] &= - \frac{1}{2} \left( \expwrt{p}{\log\left(\frac{p(x)}{p(x) + q(x)}\right)} + \expwrt{q}{\log\left(\frac{q(x)}{p(x) + q(x)}\right)} \right)
\\
&= \log(2) - \frac{1}{2}\left(\KL{p}{\tfrac{1}{2}(p+q)} + \KL{q}{\tfrac{1}{2}(p+q)}\right)
\\
&= \log(2) - \JSD{p}{q}.
\end{align*}
It is typically not possible for a machine learning algorithm to match the optimal classifier exactly. 
However, $\Lbar$ is \emph{minimized} at $\phimin$, so for \emph{any} function $\phi$,
$$
\Lbar[\phi] = \log(2) - D_{JS}(p, q) + E,
$$
where the training error, $E$, is strictly non-negative.
So regardless of what function the classifier actually converges onto,
$$
\JSD{p}{q} \geq \log(2) - \Lbar[\phi],
$$
and equality is achieved at $\phimin$. Using the sample estimator for $\Lbar$ in Equation~\eqref{eq:loss}, we appear to have attained our first technique for estimating the divergence between two distributions using only samples from $p$ and $q$,
\begin{align}
    \label{eq:jsdhat}
    \JSDhat{p}{q} = \log(2) + \frac{1}{N}\sum_{i=1}^N \left( l_i \log(\phi(x_i)) + 
    (1 - l_i) \log(1 - \phi(x_i)) \right)
\end{align}
where
\begin{align}
\expect{\JSDhat{p}{q}} \leq \JSD{p}{q}.
 \label{eq:jsdhatsecond}
\end{align}

However there is one important caveat. It is important to notice that although $\Lbar$ is minimized by the optimal classifier of $p$ and $q$, the sample estimator $\Lhat$ in Equation~\eqref{eq:loss} is not. 
The sample loss is in fact minimized by the function $\phi(x) = \frac{N_p(x)}{N_p(x) + N_q(x)}$ where $N_p(x)$ and $N_q(x)$ give the number of times the value $x$ is encountered in the training dataset associated with categories $p$ and $q$ respectively.
In other words, the sample loss is minimized by the optimal classifier of the two sample-distributions, not the optimal classier of the population distributions.
Similarly, $\Lhat[\phi]$ is not an unbiased estimator for $\Lbar[\phi]$ when evaluated on the dataset used to train the classifier.
So, just like any other machine learning problem, over-fitting is a concern.
Fortunately, when evaluated on an independent set of validation data-points, the relation
$$
\expect{\Lhat_\mathrm{val}[\phi]} = \Lbar[\phi]
$$
holds, inequality~\eqref{eq:jsdhatsecond} is satisfied, and the expected value of 
 the estimator $\JSDhat{p}{q}$ evaluated on the validation set provides a reliable lower bound of the true value of $\JSD{p}{q}$.
As with any estimator, $\JSDhat{p}{q}$ has an associated uncertainty which may be estimated from the data. If one assumes that a fixed number of samples from each category, $N_\mathrm{val}$, are used for validation, then Equation \eqref{eq:jsdhat} reduces to
\begin{align}
    \JSDhat{p}{q} = \log (2)+\frac{1}{2}\left(\frac{1}{N_\mathrm{val}} \sum_{x \text{ from } p} \log \left(\phi\left(x\right)\right)+\frac{1}{N_\mathrm{val}} \sum_{x \text{ from } q} \log \left(1-\phi\left(x\right)\right)\right),
\end{align}
and the standard sample mean uncertainties of the two sums may be combined using standard rules to give an uncertainty estimate for $\JSDhat{p}{q}$. If the number of samples in each category is allowed to vary, simple generalisations may be derived in each case.

\begin{figure}[h]
\includegraphics[width=1.0\textwidth]{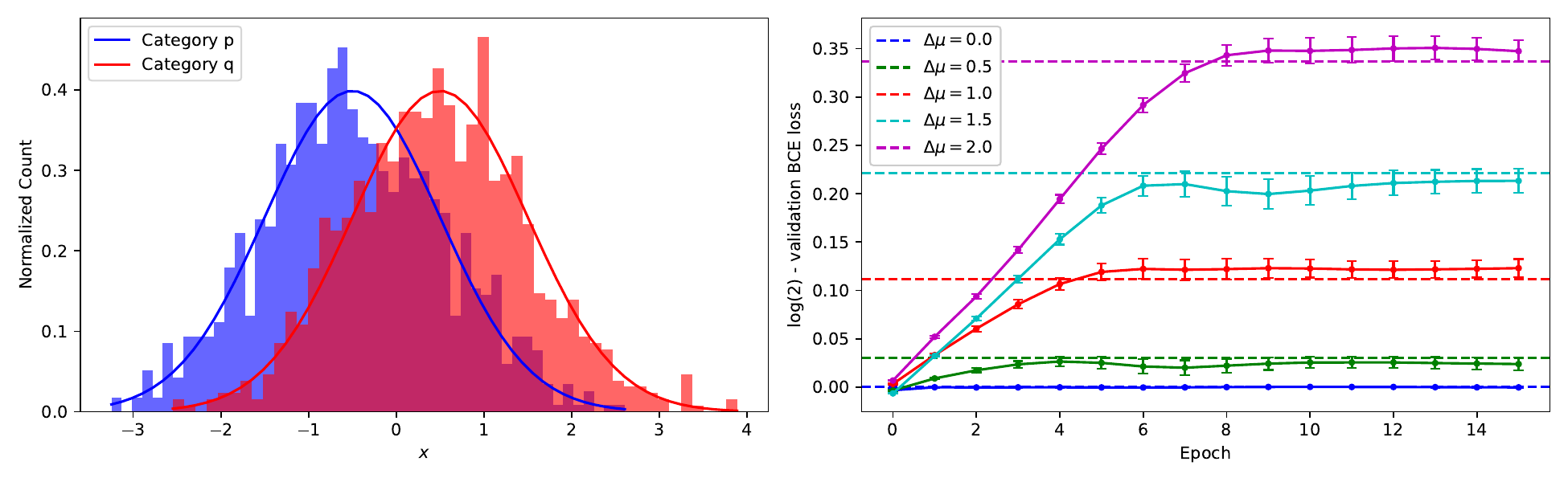}
    \caption{Left: An example of two normal distributions on the real numbers, $p$ and $q$, with unit standard deviation and \emph{separation} $\Delta\mu := \mu_q - \mu_p = 1.0$ overlaid onto two histograms of $1000$ samples from each distribution. We use $p$ and $q$ as both the category label and the density function, $p(x)$ and $q(x)$, where $x$ is used to denote an arbitrary real number. Right: Each dashed line shows the numerically integrated Jensen-Shannon divergence between $p$ and $q$ as the separation $\Delta\mu$ is varied. For each separation, the solid line shows the convergence of $\JSDhat{p}{q} = \log(2) - \Lhat_{\text{val}}$ onto $\JSD{p}{q}$ for a binary classifier trained and validated on separate datasets of 1000 samples per category.}
\label{fig:JSD}
\end{figure}

Figure \ref{fig:JSD} shows the results of the technique applied to pairs of 1D Gaussian distributed categories.
A simple neural network consisting of a sequence of linear layers ($1\times64\times64\times64\times1$) separated by LeakyReLU activation functions and ending in a sigmoid normalisation was trained on $1000$ examples from each category to optimise a binary cross entropy loss function.
The value of $\JSDhat{p}{q} = \log(2) - \Lhat_{\text{val}}$ was evaluated on an independent set of $1000$ samples from each category at the end of each epoch. The convergence of $\JSDhat{p}{q}$ onto $\JSD{p}{q}$ is shown in the right hand panel consistent with a training error of $E \approx 0$ and fluctuations consistent with the error-bars derived as previously described.

To map this example into one of hypothesis testing, consider a set of measured data points and a number of hypotheses which each predict Gaussian distributed measurements with varying means.
The results in Figure \ref{fig:JSD} demonstrate how we might easily find the hypothesis which, according to our best estimate, has the smallest divergence to the data distribution using samples from each hypothesis.
Although the example is extremely simple, the beauty of this approach is that the complexity of the algorithm effectively doesn't scale at all with the complexity of the category distributions, nor the dimension of the categories.
The same cannot be said for the binned and likelihood approximation approaches mentioned in the introduction. 

\section{Dual representations, $f$-divergences and functional optimization}
\label{sec:fdivergences}

Section \ref{sec:jsd} provides the most familiar example of a divergence calculation through machine learning that we know of, but as it so happens, a wide range of theorems exist relating all sorts of divergences to a corresponding variational problem referred to as a \emph{dual representation}.
Some authors even explore the possibility of calculating these divergences using machine learning \cite{GAN, wGAN, IPM, renyi} but we believe these miss the application for these tools in the experimental sciences.
In our work so far, we have focused on the class of \emph{$f$-divergences} which include many of the common divergences encountered in other contexts. These include the Jensen-Shannon divergence, KL divergence, the total variational distance, and many others.
Every $f$-divergence is identified by a function $f:[0,\infty) \rightarrow (-\infty,\infty]$ which must:
\begin{enumerate}
    \item be convex;
    \item satisfy $f(1) = 0$;
    \item be finite everywhere except possibly at $f(0)$; and
    \item be right continuous at $0$, that is $\lim_{t\rightarrow0^+} f(t) = f(0)$, although $f(0)$ may be infinite.
\end{enumerate}
Given such a \emph{generating} function, the $f$-divergence between a distribution $p$ absolutely continuous with respect to a distribution $q$ is defined by
\begin{align}
    \label{eq:df}
    \Df{p}{q} := \int \mathrm{d}x \, q(x) \,f\left(\frac{p(x)}{q(x)}\right) = \expwrt{q}{f\left(\frac{p(x)}{q(x)}\right)}.
\end{align}
A dual representation of a given $f$-divergence may be constructed using the Legendre transformation of $f$, denoted as $f^*$, and the trick \cite{Polyanskiy}
\begin{align}
    \Df{p}{q} = \sup_{\phi \in \mathcal{F}} \left\{ \expwrt{p}{\phi(x)} - \expwrt{q}{f^*(\phi(x))} \right\}.
     \label{eq:dfdual-basic}
\end{align}
The supremum is taken over the set of all functions, $\mathcal{F}$, from the sample space $\Omega$ to the domain of $f^*$.
Using the well-known properties of the Legendre transform and functional variation of the supremund in Equation~\eqref{eq:dfdual-basic}, one may derive a convenient alternative form for any generating function differentiable on $(0, \infty)$,
\begin{align}
    \Df{p}{q} = \sup_{\phi: \Omega \rightarrow \mathbb{R}} \left\{ \expwrt{p}{f'\left(e^{\phi(x)}\right)} - \expwrt{q}{f^*\left(f'\left(e^{\phi(x)}\right)\right)} \right\}.
     \label{eq:dfdual}
\end{align}
In this case we are free to take the supremum over all functions from $\Omega$ to the whole real number line, and the supremum is attained by the log-likelihood ratio $\phi(x)=\log\left(\frac{p(x)}{q(x)}\right)$.
Fortunately the majority of generating functions of interest fit into this category.
Table~\ref{tbl:fdivergences} provides some examples and the functions needed to implement Equation~\ref{eq:dfdual-basic}~or~\ref{eq:dfdual}.
\begin{table}[h]
\begin{tabular}{ c | c | c | c | c }
 \hline
 Name & Generating function, $f$ & $f^*$ & $\operatorname{dom}(f^*)$ & $f'$ \\ [0.5ex] 
 \hline\hline
 KL & $t \ln t$ & $e^{t-1}$ & $\mathbb{R}$ & $1+\log(t)$ \\ 
 Jensen-Shannon & $\frac{1}{2}\left(t \ln t-(t+1) \ln \left(\frac{t+1}{2}\right)\right)$ & $-\frac{1}{2}\ln(2-e^{2t})$ & $\left(-\infty, \tfrac{1}{2}\ln(2)\right)$ & $\frac{1}{2} \log\left( \frac{2t}{t + 1} \right)$ \\ 
 Total variational & $\frac{1}{2}|t-1|$ & $t$ & $\left[-\frac{1}{2}, \frac{1}{2}\right]$ & N/A \\  
 $\chi^2$ & $\frac{1}{2}(t-1)^2$ & $t\left(\tfrac{1}{2}t+1\right)$ & $\mathbb{R}$ & $t-1$
\end{tabular}
\caption{A few examples of common $f$-divergences, their generating functions, corresponding Legendre transforms, Legendre transform domains, and generating function first derivatives.}
\label{tbl:fdivergences}
\end{table}

Using Equation~\eqref{eq:dfdual} it is simple to construct an estimator for a lower bound of $\Df{p}{q}$ in almost complete analogy with the estimator $\JSDhat{p}{q}$ in the previous section, since the expectation values over $p$ and $q$ may be estimated using samples from $p$ and $q$.
Picking the KL-divergence as an example, given two datasets sampled from distributions $p$ and $q$, we may lower bound $\KL{p}{q}$ by following these steps:
\begin{enumerate}
    \item Partition each of the two datasets into a training and validation set.
    \item Use batched gradient descent to maximise the functional
    $$
    \KLhat{p}{q} = \frac{1}{N_p} \sum_{x \text{ from } p} \left(1 + \phi(x)\right) - \frac{1}{N_q} \sum_{x \text{ from } q} e^{\phi(x)}
    $$
    on the training dataset for some a machine learning model $\phi: \Omega \rightarrow \mathbb{R}$. $N_p$ and $N_q$ denote the number of samples drawn from distributions $p$ and $q$ in the given batch.
    \item At the end of each epoch evaluate $\KLhat{p}{q}$ on the whole validation dataset, unbatched, to obtain an estimate, along with error-bars, for a lower bound of $\KL{p}{q}$.
    \item Repeat steps 2 \& 3 until successive lower bounds stop increasing.
\end{enumerate}

Having trained $\phi(x)$ to provide an approximation of the log-likelihood ratio, one may use its output to lower bound the value of any other $f$-divergence generated by a second differentiable function $f_2$ by evaluating the sample estimator of Equation~\ref{eq:dfdual} on the validation dataset using $f_2^*$ and $f_2'$.
This interesting property enables the possibility of reporting a different $f$-divergence to the one used to train the network.
It is worth pointing out that the dual representation presented in Equation \eqref{eq:dfdual-basic} is not unique, and is sometimes referred to as the \emph{naive variational representation} of an $f$-divergence. 
A number of improved dual representations of $f$-divergences have been proposed \cite{Polyanskiy}, but we have not yet fully explored these representations, and are considered out of scope for this paper.
Future work aims to investigate whether the dual representations of certain $f$-divergences have better convergence properties than others, and whether these may be used to train $\phi$ and then construct better estimates of the $f$-divergences we choose to report.

Although divergences which are not $f$-divergences crop up in the machine learning literature, Wasserstein GANs for example \cite{wGAN}, $f$-divergences have a number of properties which naturally lend themselves to scientific disciplines.
These advantages include co-ordinate independence, and the data processing inequality, two prerequisites for any reasonable measure between two distributions of physical quantities.

\section{An example calculating the KL-divergence between two high dimensional distributions}
\label{sec:paritydatasection}

Having established the machinery to estimate a number of divergences, we move on to a non-trivial toy problem that demonstrates some of the advantages and quirks of this approach in practice.
The aim of this section is not to solve the actual problem presented, but to make certain points about the techniques described, so the actual machine learning models used will remain simple and intentionally unoptimized.



\begin{figure}[h]
\includegraphics[width=1.0\textwidth]{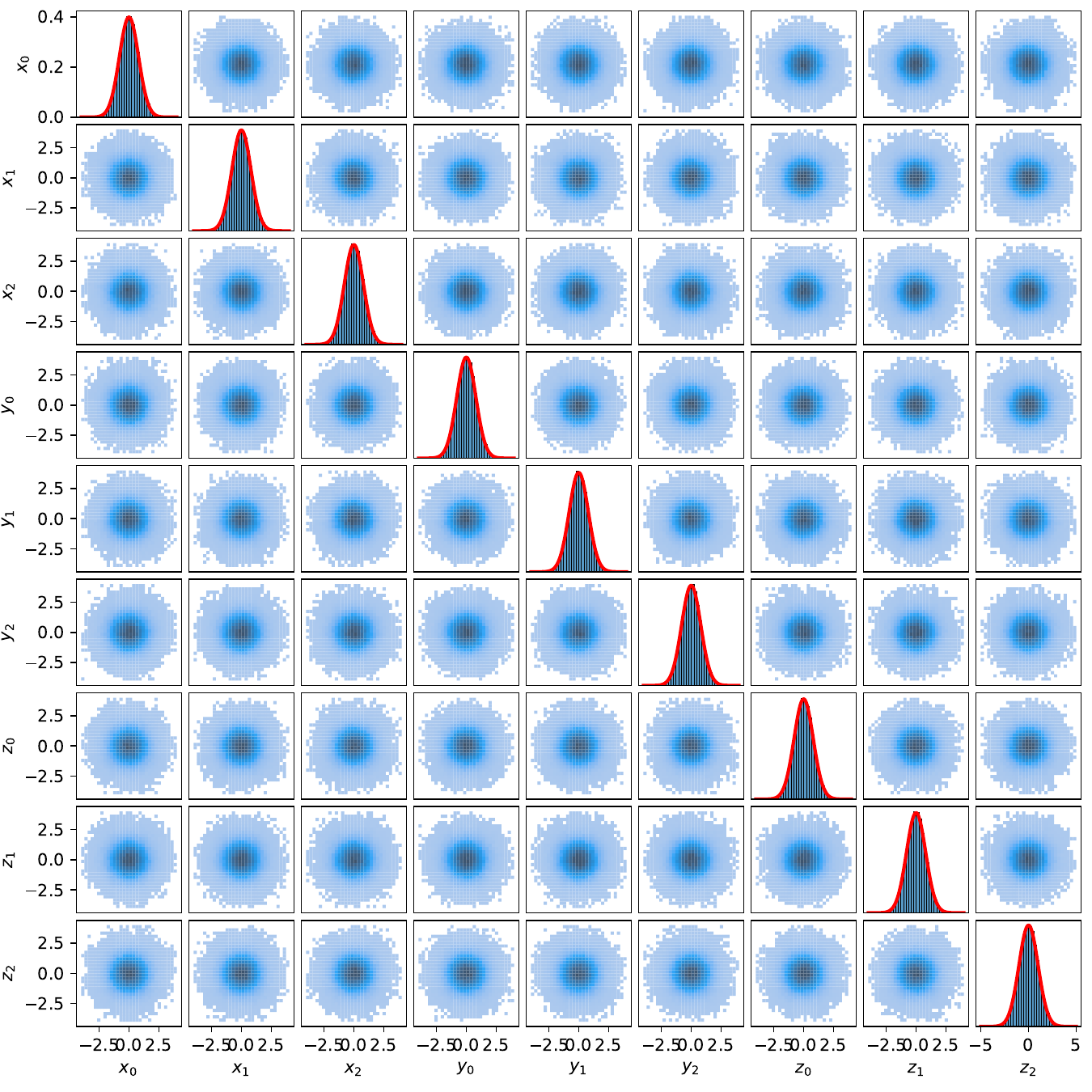}
    \caption{\textbf{On the diagonal:} A 1D histogram of each of the nine components which define the three 3D vectors - $\vec{x},\vec{y},\vec{z}$ - in the dataset. The unit normal distributions superimposed in red demonstrate that each component's marginal distribution is unit-normal distributed. \textbf{On the off-diagonal:} 2D histograms of every pair of components in the dataset. These demonstrate that once marginalized over the remaining 7 components, the distributions of the remaining 2 coordinates are independent of one another.}
\label{fig:parity_pairplot}
\end{figure}

\begin{figure}[h]
\includegraphics[width=1.0\textwidth]{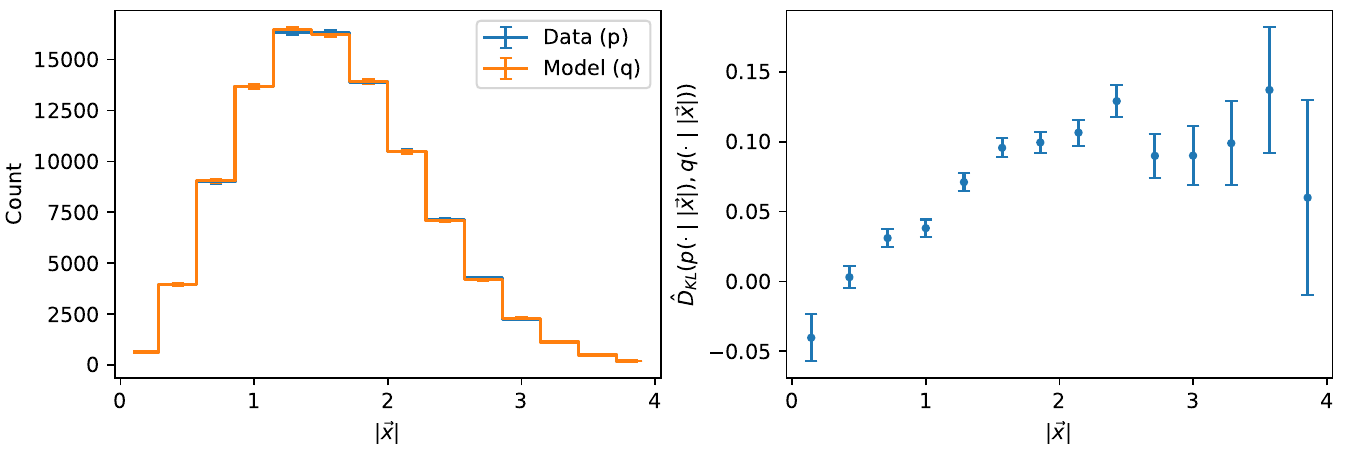}
    \caption{\textbf{Left:} A histogram of the magnitude of the first vector, $|\vec{x}|$, in the original data and the straw model. These histograms suggest there is no difference in the distribution of the first vector's magnitude between the two. \textbf{Right:} Each point indicates a lower bound on the KL divergence between the model and data distributions conditioned on a given $|\vec{x}|$ bin. The machine learning model is evidently better at separating the two datasets at larger values of $|\vec{x}|$ which provides a clue on the nature of the difference between the datasets. These lower bounds were obtained from the machine learning model used to lower bound the unconditioned KL divergence, without re-training the model.}
\label{fig:mag_x_hist}
\end{figure}

\begin{figure}[h]
\includegraphics[width=1.0\textwidth]{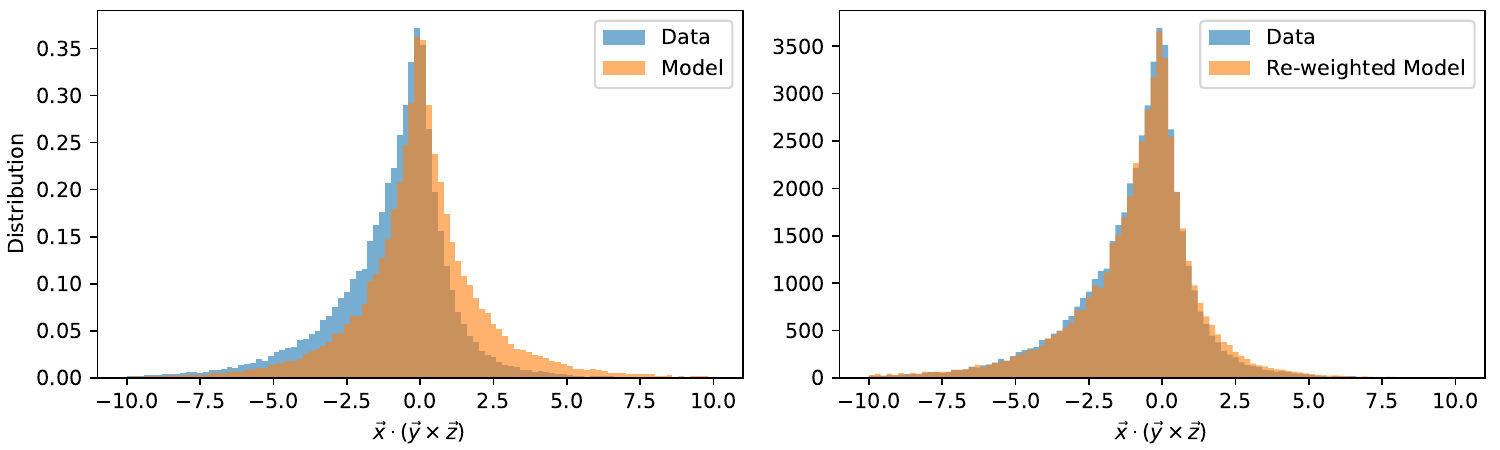}
    \caption{\textbf{Left:} The distribution of the parity variable, $P := \vec{x} \cdot \left(\vec{y}\times\vec{z}\right)$, of the original data and the straw model. \textbf{Right:} The same the parity variable histograms, however, the straw model distribution has been re-weighted with the likelihood ratio learnt by the machine learning model in the process of calculating the KL divergence between the distributions. Although the re-weighted distribution is not in perfect agreement with the original data, this plot demonstrates that the model has learnt almost all the differences between the distributions contained in this variable.}
\label{fig:parity_reweight}
\end{figure}

Consider a set of data points corresponding to the measurement of triplets of 3D vectors drawn from some unknown distribution $p$. 
For now, the source of these vectors is not important, but one can imagine they represent the momentum vectors of three particles produced in an atomic decays, or the arrangement of triplets of galaxies relative to the earth.
Suffice to say we are talking about a list of 100000 samples of three 3D vectors per sample.
Figure~\ref{fig:parity_pairplot} shows various 1D and 2D component-wise histograms of the data which suggest that the distribution of every component is an independent standard normal distribution; a reasonable first guess.
This hypothesis is extremely easy to simulate and is guaranteed to reproduce every one of the histograms in Figure~\ref{fig:parity_pairplot}, up to statistical fluctuations.
But how can we be sure that there aren't features hiding in the data which simply don't appear in the histograms we've thought to check so far?
Using our machine learning approach to lower bound a divergence between the distributions provides such a global check which is maximally sensitive, provided the network is given the data and conditions needed to converge.

Putting Equation~\eqref{eq:dfdual} into practice, a dense network ($9\times 128 \times 64 \times 64 \times 64 \times 64 \times 1$, with LeakyReLU activation functions) was trained as described in Section~\ref{sec:fdivergences} with a 50-50 train/validation split to estimate the KL divergence between the data and a competing `straw model' which assumes independent unit normal distributed components for each vector, as guessed above.
We use the letter $q$ to denote the underlying distribution of the straw model.
Once trained, this simple setup demonstrated that the KL divergence between the underlying data distribution and our straw model is at least $0.077 \pm 0.003$ in just a couple minutes.
This constitutes overwhelming evidence that there are features in the data - not visible in the histograms we bothered to check - that are missing in the straw model.
If this were a real problem, this would be the point where the fun would really start, knowing that interesting features exist in the data and trying to understand what the machine learning algorithm has found.

A useful tool in such a search is to study the behaviour of the divergence as a function of some variable in the data, the magnitude of the first vector, $|\vec{x}|$, for example, by constructing a lower bound on the divergence between the two underlying distributions conditioned on a particular value of $|\vec{x}|$, $\KL{p(\cdot \mid |\vec{x}|)}{q(\cdot \mid |\vec{x}|)}$.
This can be achieved by binning the data based on the variable of interest, and evaluating $\KLhat{p}{q}$ on the samples within each bin.
However one should not use the network trained on the full dataset directly as for a given sample, $w$, the network's output provides an approximation of $\log\left(\frac{p(w)}{q(w)}\right)$ whereas $\hat{D}_\text{KL}$ evaluated on the samples in a particular bin of $|\vec{x}|$ is maximised by $\log\left(\frac{p(w|\mid\vec{x}|)}{q(w\mid|\vec{x}|)}\right)$.
One can adjust the output of the network by looping through the training dataset a final time and constructing a histogram of the two datasets, as shown in the left panel of Figure~\ref{fig:mag_x_hist}.
Then using the relationship $\frac{p(w|\mid\vec{x}|)}{q(w\mid|\vec{x}|)} = \frac{p(w)}{q(w)} \frac{q(|\vec{x}|)}{p(|\vec{x}|)}$, create an estimator for $\frac{p(w|\mid\vec{x}|)}{q(w\mid|\vec{x}|)}$ by re-weighting the networks output by the ratio of the two histograms within each bin.
In our particular case, this reweighting has almost no effect as the distributions are almost identical in $|\vec{x}|$, but this is not the case in general.
The right panel of Figure~\ref{fig:mag_x_hist} shows what this looks like in practice for our example and suggests that the difference between the original data and the straw model increases as $|\vec{x}|$ increases.
Although we have some more ideas about how searches can be done in practice, this is out of scope for this paper and lots of work exists in the literature on machine learning interpretability which may be used.
Since we know exactly how this dataset was produced, we use the opportunity to instead make some other points.

The original data differs from the straw model in its distribution of the so-called \emph{parity} defined by,
$$
P := \vec{x} \cdot \left(\vec{y} \times \vec{z}\right).
$$
The parity distribution for the original data and the straw model is shown in the left panel of Figure~\ref{fig:parity_reweight}.
The straw model predicts a symmetric parity distribution, whereas the actual data is skewed towards negative parity.
Since $\phi(x)$ provides an estimate of the log-likelihood ratio, we can check that the network has learnt to exploit this difference in $P$ by comparing the histogram of $P$ in the original dataset to the histogram of $P$ in the independent model with each sample, $x$, weighted by the network's estimate of the likelihood ratio $e^{\phi(x)} = \frac{p(x)}{q(x)}$.
If $\phi$ is well trained these histograms will coincide.
The right panel of Figure~\ref{fig:parity_reweight} shows that our example network has learnt the differences in the parity distributions quite well even without any attempts to fine-tune the machine learning.

\begin{figure}[h]
\centering
\includegraphics[width=1.0\textwidth]{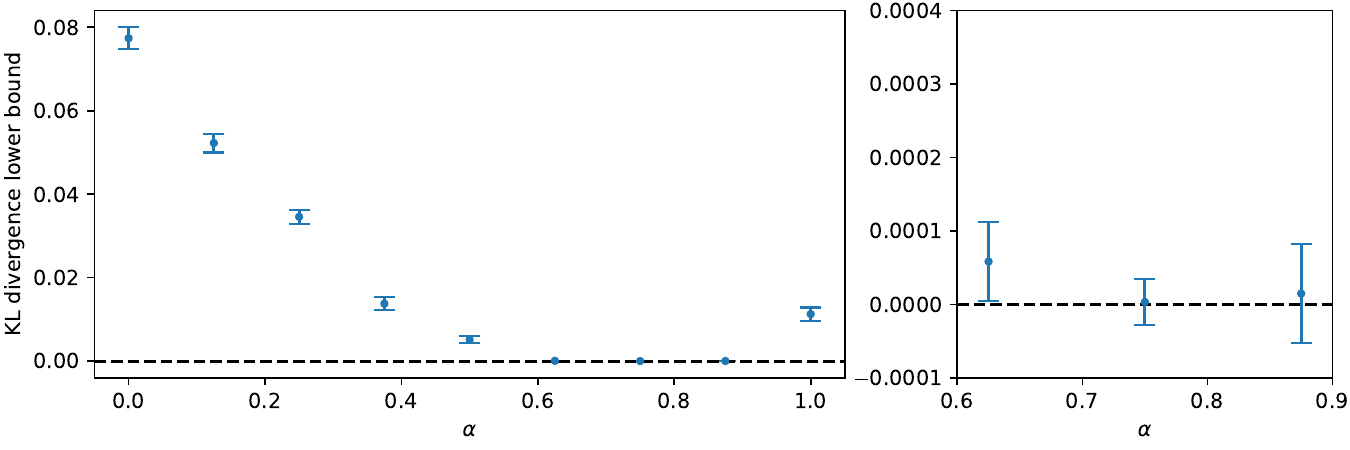}
\caption{\textbf{Left:} The lower bound of the KL-divergence obtained as a function of the asymmetry parameter used to generate the model data. \textbf{Right:} Same but zoomed into the range $(0.6, 0.9)$. The value $\alpha=0.75$ was used to produce the original data.}
\label{fig:KL_versus_alpha}
\end{figure}

Appendix~\ref{apx:parity} explains how the data was produced, and that the degree of asymmetry in the parity of the data is controlled by an additional parameter $\alpha \in [0, 1]$. The value $\alpha=0$ is equivalent to the straw model, and $\alpha=1$ corresponds to a large degree of asymmetry.
The training procedure was repeated for a range of values of $\alpha$ and the results are shown in Figure~\ref{fig:KL_versus_alpha}.
These results can be summarised as follows:

\say{We observe that the KL-divergence lower bound attained by the procedure decreases until it is consistent with $0$ in an interval around $0.625-0.875$.
This is consistent with the value of $\alpha=0.75$ used to generate the original data.
However, with the amount of data given, this particular machine learning model was not able to exclude the values of $\alpha=0.625$ nor $0.875$.}

\subsection{Discussion}

The phrasing of the results in the preceeding paragraph were carefully chosen since strictly speaking this technique can only ever lower-bound the true divergence.
In simple terms, obtaining a lower bound for the KL-divergence consistent with 0 is compatible with there being no observable differences between two datasets, however this does not constitute proof that none exists; even differences which are in principal noticeable within the given datasets.
This issue is reminiscent of the argument made in the introduction that just because two datasets agree on a particular histogram you decided to check, does not mean that the two datasets are indistinguishable.
The difference here is that our approach is guided by gradient descent and remains globally sensitive to all possible differences in the data, as opposed to the established techniques which are restricted to our inspired guesses of what variables to check.

It is worth highlighting that it is difficult to dream up a relatively simple toy example which is easy to simulate, understandable by the typical reader, and yet complex enough to fully justify the use of the techniques described.
Since toy models are simulated, we always know how the original data was produced, and therefore it is very artificial to constrain oneself to an incomplete model space.
As a result, the series of models considered in Figure~\ref{fig:KL_versus_alpha} did in fact contain the `true' hypothesis, however none of the analysis relied on this fact and we would be free to make conclusions about which model performs the best had we chosen an incomplete model space in our example.
Interestingly, if one does end up working in an incomplete model space, the methods described are capable of alerting us to this fact if the divergence lower bound obtained for all models is greater than $0$ with statistical-significance.
This is something which a typical Bayesian or binary hypothesis test cannot establish.

We also hope it is clear that even though the class of models described in Appendix~\ref{apx:parity} admit a simple to evaluate likelihood function, this was neither needed nor used by any part of the statistical analysis.
Furthermore, it does not take much additional complexity for the full likelihoods of simple models to become uncomputable.
Accounting for the effects of a flawed measurement apparatus, for example, could introduce an angle-dependent measurement resolution of each vector's components, the possibility for vectors close to one another to be mistakenly merged into a single vector, and many other such effects which are easy to simulate as a Markov chain, but result in tough to compute likelihood functions.
In this case the advantages of a machine learning approach are even more pertinent.

\section{Conclusion and Outlook}

This paper was written to point out how modern techniques for estimating statistical divergences provide a globally sensitive approach to data analysis uniquely suitable for experiments producing data with ever increasing dimension and complexity.
The authors' own field of particle physics is notoriously cursed with scouring a high dimensional space for any deviations away from the stubbornly successful standard model of particle physics.
We are in the early stages of applying these techniques in a search for evidence of new physics, building on top of the work described here \cite{Lester2022, lester2022using, Tombs_2022,Birman:2022xzu}. In addition to searching for new physics, we suggest divergences could be used to benchmark and quantifying the performance of various Monte-Carlo generators, which are extremely difficult to compare as the data they produce is high dimensional and small changes can have effects in many places.
No doubt these applications will raise practical issues which will need to be addressed. A number have already become apparent, for example:

\begin{enumerate}
    \item A machine learning model is free to pick up on any differences between data and whatever models we propose, including uninteresting effects due to detector mismodelling. How can one learn about how the machine learning model is differentiating the datasets and update one's detector model if required. Significant effort has been invested towards this point and we intend to write up our results in the coming months.
    \item What is the best way to split up train and validation data? More training data results in a higher expected lower-bound on the divergence of choice, but less validation data results in a lower confidence in the value of the lower bound obtained as reflected by larger error bars. This is particularly important when you believe the differences between two underlying distributions is extremely small. Empirical studies on the scaling of network performance as a function of the amount of training data are interesting, but it is unclear how these trends might generalise to arbitrary machine learning problems \cite{Hestness2017DeepLS}.
    \item What techniques can one use to validate the convergence of the machine learning model and therefore the quality of the divergence lower-bound obtained. If one obtains a lower bound of the divergence between data and two models A\&B, under what conditions - if any - can we reliably compare the two lower bounds to conclude which model is performing better? The aforementioned empirical studies are once again relevant to understanding the degree of training error, but once again do not generalise \cite{Hestness2017DeepLS}.
    \item In the context of $f$-divergences, what is the best way to train the machine learning model and which $f$-divergence should one report? In a complete model space, all divergences must agree on which model is performing best\footnote{This follows from $\D{p}{q} = 0 \iff p = q$.}, but in an incomplete model space various divergences may disagree on which model is `closest' to the data. A natural choice is to report the KL-divergence between the data and each model, since the conclusions obtained should align with those obtained via traditional log-likelihood ratio techniques in the large data limit. However, the KL divergence is unbounded and $\KL{p}{q}$ is only defined if $q(x)=0 \implies p(x)=0$. In contrast the definition of some $f$-divergences, like the Jensen-Shannon divergence, are bounded and can be extended to compare any two distributions on the same space.
\end{enumerate}
We hope to inspire a number of readers to try apply these techniques within their own fields and to help settle some of these unanswered questions.

\begin{appendices}

\section{How the parity data was produced} \label{apx:parity}

The asymmetric data used and described in Section~\ref{sec:paritydatasection} was produced by first sampling each component from a unit normal distribution, just like in the independent-component `straw' model.
Then, each sample was either flipped, or not, under the transformation
\begin{align*}
    \vec{x} \rightarrow -\vec{x} \phantom{,}\\ 
    \vec{y} \rightarrow -\vec{y} \phantom{,}\\ 
    \vec{z} \rightarrow -\vec{z},
\end{align*}
which has the effect of sending $P \rightarrow -P$.
The probability, $g$, of flipping a given sample was a function of the sample's original parity, $P$.  Specifically $g$ was chosen to be $\alpha S(10 P)$, where $S$ is the sigmoid function, and $\alpha$ is a parameter which sets the total degree of asymmetry.
Thus, samples with $P \gg 0$ were flipped at a rate of $\alpha$, and samples with $ P \ll 0 $ were almost never flipped. Hence the heavier tails towards negative $P$.
Augmentation of the vectors with this procedure leaves the marginal distributions over single coordinates and pairs of coordinates unchanged, which is why the flipping is not evident in Figure~\ref{fig:parity_pairplot}. The data discussed in Section~\ref{sec:paritydatasection} was produced with $\alpha=0.75$.

\section{Code which supports this work}
A python package named ``iwpc''\footnote{The letters in ``ipwc'' stand for ``I will prove convergence''.} \cite{IWPC:pypi} has been made available on the Python Package Index (PyPI, \cite{pypi}) to permit others to evaluate divergences on samples from their own distributions using the methods described in this paper. The underlying software packaged within ``iwpc'' may be found in a separate development repository \cite{IWPC:bitbucket}.
A number of examples can be found in the examples directory including code to reproduce the results of the parity example in this paper.
Those who make use of \cite{IWPC:pypi} or \cite{IWPC:bitbucket} in their own work are asked to consider citing the present paper. 

\end{appendices}

\bibliography{sn-bibliography}


\begin{thebibliography}{16}
\ifx \bisbn   \undefined \def \bisbn  #1{ISBN #1}\fi
\ifx \binits  \undefined \def \binits#1{#1}\fi
\ifx \bauthor  \undefined \def \bauthor#1{#1}\fi
\ifx \batitle  \undefined \def \batitle#1{#1}\fi
\ifx \bjtitle  \undefined \def \bjtitle#1{#1}\fi
\ifx \bvolume  \undefined \def \bvolume#1{\textbf{#1}}\fi
\ifx \byear  \undefined \def \byear#1{#1}\fi
\ifx \bissue  \undefined \def \bissue#1{#1}\fi
\ifx \bfpage  \undefined \def \bfpage#1{#1}\fi
\ifx \blpage  \undefined \def \blpage #1{#1}\fi
\ifx \burl  \undefined \def \burl#1{\textsf{#1}}\fi
\ifx \doiurl  \undefined \def \doiurl#1{\url{https://doi.org/#1}}\fi
\ifx \betal  \undefined \def \betal{\textit{et al.}}\fi
\ifx \binstitute  \undefined \def \binstitute#1{#1}\fi
\ifx \binstitutionaled  \undefined \def \binstitutionaled#1{#1}\fi
\ifx \bctitle  \undefined \def \bctitle#1{#1}\fi
\ifx \beditor  \undefined \def \beditor#1{#1}\fi
\ifx \bpublisher  \undefined \def \bpublisher#1{#1}\fi
\ifx \bbtitle  \undefined \def \bbtitle#1{#1}\fi
\ifx \bedition  \undefined \def \bedition#1{#1}\fi
\ifx \bseriesno  \undefined \def \bseriesno#1{#1}\fi
\ifx \blocation  \undefined \def \blocation#1{#1}\fi
\ifx \bsertitle  \undefined \def \bsertitle#1{#1}\fi
\ifx \bsnm \undefined \def \bsnm#1{#1}\fi
\ifx \bsuffix \undefined \def \bsuffix#1{#1}\fi
\ifx \bparticle \undefined \def \bparticle#1{#1}\fi
\ifx \barticle \undefined \def \barticle#1{#1}\fi
\bibcommenthead
\ifx \bconfdate \undefined \def \bconfdate #1{#1}\fi
\ifx \botherref \undefined \def \botherref #1{#1}\fi
\ifx \url \undefined \def \url#1{\textsf{#1}}\fi
\ifx \bchapter \undefined \def \bchapter#1{#1}\fi
\ifx \bbook \undefined \def \bbook#1{#1}\fi
\ifx \bcomment \undefined \def \bcomment#1{#1}\fi
\ifx \oauthor \undefined \def \oauthor#1{#1}\fi
\ifx \citeauthoryear \undefined \def \citeauthoryear#1{#1}\fi
\ifx \endbibitem  \undefined \def \endbibitem {}\fi
\ifx \bconflocation  \undefined \def \bconflocation#1{#1}\fi
\ifx \arxivurl  \undefined \def \arxivurl#1{\textsf{#1}}\fi
\csname PreBibitemsHook\endcsname

\bibitem[\protect\citeauthoryear{Cowan et~al.}{2011}]{Cowan2011}
\begin{botherref}
\oauthor{\bsnm{Cowan}, \binits{G.}},
\oauthor{\bsnm{Cranmer}, \binits{K.}},
\oauthor{\bsnm{Gross}, \binits{E.}},
\oauthor{\bsnm{Vitells}, \binits{O.}}:
Asymptotic formulae for likelihood-based tests of new physics.
The European Physical Journal C
\textbf{71}(2)
(2011)
\doiurl{10.1140/epjc/s10052-011-1554-0}
\end{botherref}
\endbibitem

\bibitem[\protect\citeauthoryear{Aaij et~al.}{2022}]{LHCB}
\begin{barticle}
\bauthor{\bsnm{Aaij}, \binits{R.}}, \betal:
\batitle{{Measurement of the $B^0_s\to\mu^+\mu^-$ decay properties and search
  for the $B^0\to\mu^+\mu^-$ and $B^0_s\to\mu^+\mu^-\gamma$ decays}}.
\bjtitle{Phys. Rev. D}
\bvolume{105}(\bissue{1}),
\bfpage{012010}
(\byear{2022})
\doiurl{10.1103/PhysRevD.105.012010}
{\href{https://arxiv.org/abs/2108.09283}{{arXiv:2108.09283}}}
{[hep-ex]}
\end{barticle}
\endbibitem

\bibitem[\protect\citeauthoryear{Barter et~al.}{2018}]{Barter:2018xbc}
\begin{barticle}
\bauthor{\bsnm{Barter}, \binits{W.}},
\bauthor{\bsnm{Burr}, \binits{C.}},
\bauthor{\bsnm{Parkes}, \binits{C.}}:
\batitle{{Calculating $p$-values and their significances with the Energy Test
  for large datasets}}.
\bjtitle{JINST}
\bvolume{13}(\bissue{04}),
\bfpage{04011}
(\byear{2018})
\doiurl{10.1088/1748-0221/13/04/P04011}
{\href{https://arxiv.org/abs/1801.05222}{{arXiv:1801.05222}}}
{[physics.data-an]}
\end{barticle}
\endbibitem

\bibitem[\protect\citeauthoryear{Goodfellow et~al.}{2014}]{GAN}
\begin{bchapter}
\bauthor{\bsnm{Goodfellow}, \binits{I.}},
\bauthor{\bsnm{Pouget-Abadie}, \binits{J.}},
\bauthor{\bsnm{Mirza}, \binits{M.}},
\bauthor{\bsnm{Xu}, \binits{B.}},
\bauthor{\bsnm{Warde-Farley}, \binits{D.}},
\bauthor{\bsnm{Ozair}, \binits{S.}},
\bauthor{\bsnm{Courville}, \binits{A.}},
\bauthor{\bsnm{Bengio}, \binits{Y.}}:
\bctitle{Generative adversarial nets}.
In: \beditor{\bsnm{Ghahramani}, \binits{Z.}},
\beditor{\bsnm{Welling}, \binits{M.}},
\beditor{\bsnm{Cortes}, \binits{C.}},
\beditor{\bsnm{Lawrence}, \binits{N.}},
\beditor{\bsnm{Weinberger}, \binits{K.Q.}} (eds.)
\bbtitle{Advances in Neural Information Processing Systems},
vol. \bseriesno{27},
p. \bfpage{1}.
\bpublisher{Curran Associates, Inc.},
\blocation{San Diego, CA}
(\byear{2014}).
\burl{https://proceedings.neurips.cc/paper_files/paper/2014/file/5ca3e9b122f61f8f06494c97b1afccf3-Paper.pdf}
\end{bchapter}
\endbibitem

\bibitem[\protect\citeauthoryear{Arjovsky et~al.}{2017}]{wGAN}
\begin{botherref}
\oauthor{\bsnm{Arjovsky}, \binits{M.}},
\oauthor{\bsnm{Chintala}, \binits{S.}},
\oauthor{\bsnm{Bottou}, \binits{L.}}:
Wasserstein GAN
(2017)
\end{botherref}
\endbibitem

\bibitem[\protect\citeauthoryear{Sriperumbudur et~al.}{2010}]{IPM}
\begin{bchapter}
\bauthor{\bsnm{Sriperumbudur}, \binits{B.K.}},
\bauthor{\bsnm{Fukumizu}, \binits{K.}},
\bauthor{\bsnm{Gretton}, \binits{A.}},
\bauthor{\bsnm{Schölkopf}, \binits{B.}},
\bauthor{\bsnm{Lanckriet}, \binits{G.R.G.}}:
\bctitle{Non-parametric estimation of integral probability metrics}.
In: \bbtitle{2010 IEEE International Symposium on Information Theory},
pp. \bfpage{1428}--\blpage{1432}
(\byear{2010}).
\doiurl{10.1109/ISIT.2010.5513626}
\end{bchapter}
\endbibitem

\bibitem[\protect\citeauthoryear{Birrell et~al.}{2021}]{renyi}
\begin{barticle}
\bauthor{\bsnm{Birrell}, \binits{J.}},
\bauthor{\bsnm{Dupuis}, \binits{P.}},
\bauthor{\bsnm{Katsoulakis}, \binits{M.A.}},
\bauthor{\bsnm{Rey-Bellet}, \binits{L.}},
\bauthor{\bsnm{Wang}, \binits{J.}}:
\batitle{Variational representations and neural network estimation of rényi
  divergences}.
\bjtitle{SIAM Journal on Mathematics of Data Science}
\bvolume{3}(\bissue{4}),
\bfpage{1093}--\blpage{1116}
(\byear{2021})
\doiurl{10.1137/20M1368926}
{\href{https://arxiv.org/abs/https://doi.org/10.1137/20M1368926}{{https://doi.org/10.1137/20M1368926}}}
\end{barticle}
\endbibitem

\bibitem[\protect\citeauthoryear{Polyanskiy and Wu}{2022}]{Polyanskiy}
\begin{bbook}
\bauthor{\bsnm{Polyanskiy}, \binits{Y.}},
\bauthor{\bsnm{Wu}, \binits{Y.}}:
\bbtitle{Information Theory: From Coding to Learning}.
\bpublisher{Cambridge University Press},
\blocation{Cambridge}
(\byear{2022})
\end{bbook}
\endbibitem

\bibitem[\protect\citeauthoryear{Lester et~al.}{2022}]{Lester2022}
\begin{botherref}
\oauthor{\bsnm{Lester}, \binits{C.G.}},
\oauthor{\bsnm{Mastandrea}, \binits{R.}},
\oauthor{\bsnm{Noel}, \binits{D.}},
\oauthor{\bsnm{Tombs}, \binits{R.}}:
Hunting for vampires and other unlikely forms of parity violation at the large
  hadron collider.
Journal of High Energy Physics
\textbf{2022}(8)
(2022)
\doiurl{10.1007/jhep08(2022)231}
\end{botherref}
\endbibitem

\bibitem[\protect\citeauthoryear{Lester}{2022}]{lester2022using}
\begin{botherref}
\oauthor{\bsnm{Lester}, \binits{C.G.}}:
Using unsupervised learning to detect broken symmetries, with relevance to
  searches for parity violation in nature.
Transactions on Machine Learning Research
(2022)
\end{botherref}
\endbibitem

\bibitem[\protect\citeauthoryear{Tombs and Lester}{2022}]{Tombs_2022}
\begin{barticle}
\bauthor{\bsnm{Tombs}, \binits{R.}},
\bauthor{\bsnm{Lester}, \binits{C.G.}}:
\batitle{A method to challenge symmetries in data with self-supervised
  learning}.
\bjtitle{Journal of Instrumentation}
\bvolume{17}(\bissue{08}),
\bfpage{08024}
(\byear{2022})
\doiurl{10.1088/1748-0221/17/08/P08024}
\end{barticle}
\endbibitem

\bibitem[\protect\citeauthoryear{Birman et~al.}{2022}]{Birman:2022xzu}
\begin{barticle}
\bauthor{\bsnm{Birman}, \binits{M.}},
\bauthor{\bsnm{Nachman}, \binits{B.}},
\bauthor{\bsnm{Sebbah}, \binits{R.}},
\bauthor{\bsnm{Sela}, \binits{G.}},
\bauthor{\bsnm{Turetz}, \binits{O.}},
\bauthor{\bsnm{Bressler}, \binits{S.}}:
\batitle{{Data-directed search for new physics based on symmetries of the SM}}.
\bjtitle{Eur. Phys. J. C}
\bvolume{82}(\bissue{6}),
\bfpage{508}
(\byear{2022})
\doiurl{10.1140/epjc/s10052-022-10454-2}
{\href{https://arxiv.org/abs/2203.07529}{{arXiv:2203.07529}}}
{[hep-ph]}
\end{barticle}
\endbibitem

\bibitem[\protect\citeauthoryear{Hestness et~al.}{2017}]{Hestness2017DeepLS}
\begin{botherref}
\oauthor{\bsnm{Hestness}, \binits{J.}},
\oauthor{\bsnm{Narang}, \binits{S.}},
\oauthor{\bsnm{Ardalani}, \binits{N.}},
\oauthor{\bsnm{Diamos}, \binits{G.F.}},
\oauthor{\bsnm{Jun}, \binits{H.}},
\oauthor{\bsnm{Kianinejad}, \binits{H.}},
\oauthor{\bsnm{Patwary}, \binits{M.M.A.}},
\oauthor{\bsnm{Yang}, \binits{Y.}},
\oauthor{\bsnm{Zhou}, \binits{Y.}}:
Deep learning scaling is predictable, empirically.
ArXiv
\textbf{abs/1712.00409}
(2017)
\end{botherref}
\endbibitem

\bibitem[\protect\citeauthoryear{}{}]{IWPC:pypi}
\begin{botherref}
\url{https://pypi.org/project/iwpc}
\end{botherref}
\endbibitem

\bibitem[\protect\citeauthoryear{}{}]{pypi}
\begin{botherref}
Python Package Index - PyPI.
\url{https://pypi.org/}
Accessed 2021-03-28
\end{botherref}
\endbibitem

\bibitem[\protect\citeauthoryear{}{}]{IWPC:bitbucket}
\begin{botherref}
\url{https://bitbucket.org/jjhw3/divergences}
\end{botherref}
\endbibitem

\end{thebibliography}

\end{document}